\documentclass[aps,prl,twocolumn,showpacs,10pt,superscriptaddress,preprintnumbers]{revtex4-1}
\usepackage{graphicx}
\usepackage{amsmath,bm,braket}
\usepackage{color}

\newcommand{\hs}{\hat{s}}
\newcommand{\als}{\alpha_s}
\newcommand{\ep}{\epsilon}
\newcommand{\nn}{\nonumber}

\newcommand{\lp}{L_\perp}
\newcommand{\Li}{\mathrm{Li}}

\begin{document}

\title{Transverse-momentum resummation for top-quark pairs at hadron
  colliders}
\author{Hua Xing Zhu}
\email{huaxingzhu@gmail.com}
\affiliation{Department of Physics and State Key Laboratory of Nuclear Physics and Technology, Peking University, Beijing 100871, China}
\author{Chong Sheng Li}
\email{csli@pku.edu.cn}
\affiliation{Department of Physics and State Key Laboratory of Nuclear Physics and Technology, Peking University, Beijing 100871, China}
\affiliation{Center for High Energy Physics, Peking University, Beijing 100871, China}
\author{Hai Tao Li}
\author{Ding Yu Shao}
\affiliation{Department of Physics and State Key Laboratory of Nuclear Physics and Technology, Peking University, Beijing 100871, China}
\author{Li Lin Yang}
\email{yanglilin@pku.edu.cn}
\affiliation{Department of Physics and State Key Laboratory of Nuclear Physics and Technology, Peking University, Beijing 100871, China}
\affiliation{Institute for Theoretical Physics, University of Z\"urich, CH-8057 Z\"urich, Switzerland}

\begin{abstract}
  We develop a framework for a systematic resummation of the transverse momentum distribution of top-quark pairs produced at hadron colliders based on effective field theory. Compared to Drell-Yan and Higgs production, a novel soft function matrix is required to account for the soft gluon emissions from the final states. We calculate this soft function at the next-to-leading order, and perform the resummation at the next-to-next-to-leading logarithmic accuracy. We compare our results with parton shower programs and with the experimental data at the Tevatron and the LHC. We also discuss the implications for the top quark charge asymmetry.
\end{abstract}

\pacs{14.65.Ha, 12.38.Cy}

\maketitle

The top quark is of special importance in the Standard Model (SM). Due to its large mass, it couples strongly to the Higgs boson, and is crucial to the hierarchy problem. New physics (NP) models aiming at solving the hierarchy problem often predict top partners which exhibit similar properties as the top quark and may decay into it. Possible new heavy resonances usually prefer to decay into top quark pairs. Therefore, studying the top quarks can on one hand help understanding the nature of electroweak symmetry breaking, and on the other hand probe NP beyond the SM.

If a heavy resonance decays into a top quark pair, the kinematics of the $t\bar{t}$ system will then carry information of the resonance. It is therefore worthwhile to study the $t\bar{t}$ pair as a whole instead of individual top quarks. One important example is the invariant mass of the $t\bar{t}$ pair, which is very sensitive to new physics contributions. Precision predictions for this distribution has been achieved in \cite{Ahrens:2010zv}. Besides the invariant mass, another important variable is the transverse momentum $q_T$ of the $t\bar{t}$ system, which has been recently measured by both the CMS and the ATLAS collaborations at the LHC \cite{TOP-11-013, Aad:2012hg}. One reason to study this distribution is that the top quark charge asymmetry exhibits intriguing dependence on $q_T$ \cite{Abazov:2011rq}. In particular, it was shown that the asymmetry can be enhanced by restricting to the small $q_T$ region \cite{afbtheo}. The top quark charge asymmetry has received much attention recently, due to the deviation from the SM observed at the Tevatron \cite{Aaltonen:2011kc, Abazov:2011rq}. Many NP models have been proposed to explain this discrepancy (see, e.g., \cite{Kamenik:2011wt} and references therein). Studying the $q_T$-dependent asymmetry will help to clarify which model is the correct one. Similar to the asymmetry, it has been shown recently \cite{Alvarez:2012uh} that vetoing the $t\bar{t}$ transverse momentum can enhance the sensitivity of the invariant mass distribution to the effects of NP which couples mainly to quarks. This finding makes the small $q_T$ region even more important.

Making precise predictions for the small $q_T$ region, however, is theoretically challenging. As is well-known in the case of Drell-Yan and Higgs production, soft and collinear gluon emissions give rise to large logarithms of the form $\ln(q_T^2/Q^2)$ at each order in perturbation theory, where $Q \gg q_T$ is a typical hard scale of the process. The fixed-order predictions are therefore not reliable in this region. For the case of Drell-Yan and Higgs, the method to deal with this problem is the so-called Collins-Soper-Sterman (CSS) formalism \cite{Collins:1984kg}, in which the large logarithms can be resummed to all orders in the strong coupling $\alpha_s$. For $t\bar{t}$ production, on the other hand, the CSS formalism can not be directly applied due to gluon emissions from the top quarks in the final state. Therefore, for observables sensitive to the small $q_T$ region in $t\bar{t}$ production, current experimental groups usually rely on parton shower (PS) programs, which only achieves resummation at the leading logarithmic (LL) level. Ref.~\cite{oldresum} attempted an next-to-leading logarithmic (NLL) resummation by modifying the CSS formalism. However, they did not consider color mixing between singlet and octet final-states, and they missed the contributions from initial-final gluon exchange.

In this Letter, we develop a framework for $q_T$ resummation in $t\bar{t}$ production based on the soft-collinear effective theory (SCET) \cite{scet}. The framework is built upon the works \cite{BecherNeubert}, which systematically resum the large logarithms to arbitrary accuracy. A novel feature of our framework is the appearance of a transverse soft function matrix, which describes color exchange among the initial state and final state particles. Using the available ingredients, we perform the resummation at the next-to-next-to-leading logarithmic (NNLL) accuracy.

We consider the process $N_1 (P_1) + N_2(P_2) \to t(p_3) + \bar{t} (p_4) + X$. We denote the transverse momentum of the $t\bar{t}$ pair as $q_T$. In the small $q_T$ region, the differential cross section can be written as
\begin{align}
  \label{eq:master}
  \frac{d^4\sigma}{dq_T^2 \, dy \, dM \, d\cos\theta} &= \frac{8\pi\beta_t}{3sM} \sum_{i=q,\bar{q},g} \sum_{a,b} \int^1_{\xi_1} \frac{dz_1}{z_1} \int^1_{\xi_2} \frac{dz_2}{z_2} \nonumber
  \\
  &\hspace{-1em} \times f_{a/N_1}(\xi_1/z_1,\mu) \, f_{b/N_2}(\xi_2/z_2,\mu)
  \\
  &\hspace{-1em} \times C_{i\bar{i} \leftarrow ab}(z_1,z_2,q_T,M,\cos\theta,m_t,\mu) \, , \nonumber
\end{align}
where $s$ is the collider energy, $M$ and $y$ are the invariant mass and the rapidity of the top-quark pair, $\theta$ is the scattering angle between $p_3$ and $P_1$  in the center of mass frame of $t\bar{t}$ pair, $\beta_t=\sqrt{1-4m^2_t/M^2}$, $\xi_{1,2}=\sqrt{\tau} e^{\pm y}$, with $\tau=(M^2+q_T^2)/s$. We also define
\begin{gather*}
  p_1 = \xi_1 P_1 \, , \quad p_2 = \xi_2 P_2 \, , \quad \hs = (p_1+p_2)^2 \, ,
  \\
  t_1 = (p_1-p_3)^2 - m_t^2 \, , \quad u_1 = (p_2-p_3)^2 - m_t^2 \, .
\end{gather*}
The resummed formula for the partonic function $C_{i\bar{i} \leftarrow ab}$ can be written as
\begin{align}
  \label{eq:c}
  &C_{i\bar{i} \leftarrow ab} (z_1,z_2,q_T,M,\cos\theta,m_t,\mu) = \frac{1}{2} \int^\infty_0 db \, b \, J_0(bq_T) \nonumber
  \\
  &\times \exp \big[ g_i(\eta_i, L_\perp, \als) \big] \, \Big[ \bar{I}_{i/a}(z_1, L_\perp,\als) \, \bar{I}_{\bar{i}/b}(z_2,L_\perp,\als) \nonumber
  \\
  &\hspace{7em} + \delta_{gi} \, \bar{I}'_{g/a}(z_1, L_\perp,\als) \, \bar{I}'_{g/b}(z_2,L_\perp,\als) \Big] \nonumber
  \\
  &\times \mathrm{Tr} \Big[ \bm{H}_{i\bar{i}}(M,\cos\theta,m_t,\mu_h,\mu) \, \bm{S}_{i\bar{i}}(\lp,M,\cos\theta,m_t,\mu) \Big] ,
\end{align}
where $\eta_i=(C_i\als/\pi)\ln(M^2/\mu^2)$ with $C_q=C_F=4/3$ and $C_g=C_A=3$, $L_\perp=\ln(b^2\mu^2/b_0^2)$ with $b_0=2e^{-\gamma_E}$, $J_0$ is the zeroth order Bessel function. $\bm{H}_{i\bar{i}}$ are the hard functions, evolved from an appropriately chosen hard scale $\mu_h$ that minimizes the logarithms in it to $\mu$. The hard functions are matrices in color space, as indicated by the boldface letter. They are the same as in threshold resummation, whose expressions and RG evolution can be found in \cite{Ahrens:2010zv}. The functions $g_i$ and $\bar{I}_{i/a}$ are related to the transverse PDFs, whose definition and explicit NLO expressions can be found in \cite{BecherNeubert}. The functions $\bar{I}'_{g/a}$ originate from the second Lorentz structure of the gluon transverse PDF \cite{Catani:2010pd}. They start at $\mathcal{O}(\alpha_s)$ and do not contribute to the NNLL accuracy. Eq.~(\ref{eq:c}) resembles the $Q_T$ resummation formulae for Drell-Yan and Higgs production in SCET \cite{BecherNeubert}, which have been proven to be equivalent to the traditional CSS formalism for certain choice of scales. Besides the matrix form of the hard functions, a major difference of our formula with respect to the Drell-Yan and Higgs cases is the appearance of the transverse soft functions $\bm{S}_{i\bar{i}}$, whose fixed order operator definition can be written as
\begin{align}
  \label{eq:soft}
  &\bm{S}_{i\bar{i}}(\lp,M,\cos\theta,m_t,\mu) = \frac{1}{d_i} \sum_{X_s} \int \frac{d\phi_t}{2\pi} \, d^2\bm{q}_\perp \, e^{i \bm{b} \cdot \bm{q}_\perp}
  \\
  &\times \braket{0 | \bm{Y}_{i}^\dagger \bm{Y}_{\bar{i}}^\dagger \bm{Y}_{t}^\dagger \bm{Y}_{\bar{t}}^\dagger | X_s} \, \delta^{(2)}(\bm{q}_\perp + \hat{P}_\perp) \, \braket{X_s | \bm{Y}_{i} \bm{Y}_{\bar{i}} \bm{Y}_{t} \bm{Y}_{\bar{t}} | 0} \, , \nonumber
\end{align}
where the operator $\hat{P}_{\perp}$ acts on the soft final state $X_s$ giving its transverse momentum, $d_{q(g)}=3(8)$, and $\bm{Y}_a$ are soft Wilson lines along the directions of partons $a=i,\bar{i},t,\bar{t}$. The angle $\phi_t$ is the azimuthal angle of the top quark in the transverse plane. (One may define a soft function which is exclusive in $\phi_t$, but the result will be much more complicated.) For the cases of Drell-Yan and Higgs production, such soft functions are equal to their tree-level values with the analytic regulator used in \cite{BecherNeubert, Becher:2011dz}. With the presence of colored particles in the final state, that property does not hold anymore, as will be shown below. For NNLL accuracy, we need the hard and soft functions as well as the transverse PDFs to NLO, and their anomalous dimensions to two loops. NLL accuracy corresponds to one order less in all functions than NNLL accuracy.

So far we have been working in the color space formalism \cite{Catani:1996vz}. For actual computations, it is more convenient to introduce a color basis, for which we adopt the one used in \cite{Ahrens:2010zv}. In this basis, the LO soft functions are given by
\begin{align}
  \label{eq:softmatrix}
  \bm{S}^{(0)}_{q\bar{q}} =
  \begin{pmatrix}
    N_c & 0
    \\
    0 & \frac{C_F}{2}
  \end{pmatrix}
  \, , \quad \bm{S}^{(0)}_{gg} =
  \begin{pmatrix}
    N_c & 0 & 0
    \\
    0 & \frac{N_c}{2} & 0
    \\
    0 & 0 & \frac{N^2_c - 4}{2 N_c}
  \end{pmatrix}
  \, ,
\end{align}
where $N_c=3$. At NLO, the bare soft functions can be written as
\begin{align}
  \bm{S}^{(1), \text{bare}}_{i\bar{i}} = \sum_{j,k} \bm{w}^{jk}_{i\bar{i}} \, I_{jk}\, ,
\end{align}
where $\bm{w}^{jk}_{i\bar{i}}$ are color matrices, whose explicit expressions can be found in \cite{Ahrens:2010zv}. The integrals $I_{jk}$ are given by
\begin{align}
  I_{jk} &= - \frac{(4\pi\mu^2)^\epsilon}{\pi^{2-\epsilon}} \int^{2\pi}_0 \frac{d\phi_t}{2\pi} \int [dk] \, \frac{v_j \cdot v_k \, e^{-i \bm{b} \cdot \bm{k}_\perp}}{v_j \cdot k \; v_k \cdot k} \, ,
\end{align}
where $[dk]=d^dk \, (2\pi) \, \delta(k^2) \, \theta(k^0)$, and $v_j$ are dimensionless vectors along the directions of momenta $p_j$, chosen as $v_1=n=(1,0,0,1)$, $v_2=\bar{n}=(1,0,0-1)$, $v_3^2=v_4^2=1$. The above integrals contain singularities which are not regularized by dimensional regularization. We therefore introduce a regularization factor $(\nu/k^+)^\alpha$ following \cite{Becher:2011dz}, where $k^+ = n \cdot k$, $\nu$ is an unphysical scale. We find that although the individual integrals contain poles in $\alpha$, these divergences cancel in the final soft function, along with the dependence on the unphysical scale $\nu$. After renormalizing the remaining divergences in $\ep$ in the $\overline{\rm MS}$ scheme,  the finite NLO soft function can be written as
\begin{align}
  \bm{S}_{i\bar{i}}^{(1)} &= 4 L_\perp \left( 2\bm{w}^{13}_{i\bar{i}}  \ln\frac{-t_1}{m_tM}  + 2\bm{w}^{23}_{i\bar{i}} \ln\frac{-u_1}{m_tM} + \bm{w}^{33}_{i\bar{i}} \right)  \nonumber
  \\
  &\hspace{-2em} - 4 \left( \bm{w}^{13}_{i\bar{i}} + \bm{w}^{23}_{i\bar{i}} \right) \Li_2 \Biggl( 1 - \frac{t_1u_1}{m_t^2M^2} \Biggr) + 4\bm{w}^{33}_{i\bar{i}} \ln\frac{t_1u_1}{m_t^2M^2} \nonumber
  \\
  &\hspace{-2em} -  2\bm{w}^{34}_{i\bar{i}} \, \frac{1+\beta_t^2}{\beta_t} \, \bigl[ L_\perp \ln x_s + f_{34} \bigr] \, ,
\end{align}
where $x_s=(1-\beta_t)/(1+\beta_t)$ and
\begin{align}
  f_{34} &=  - \Li_2 \left( -x_s \tan^2\frac{\theta}{2} \right) + \Li_2 \left( -\frac{1}{x_s} \tan^2\frac{\theta}{2} \right) \nonumber
  \\
  &+ 4\ln x_s \ln\cos\frac{\theta}{2} \, .
\end{align}
For a consistency check, one can verify that close to the production threshold $\beta_t\to 0$, the NLO corrections to the soft functions vanish for top-quark pair in the color-singlet state, {\it i.e.}, the $(1,1)$ components of the $\bm{S}^{(1)}_{i\bar{i}}$ matrices. The reason is that near threshold soft gluons can not be emitted from color-singlet top-quark pairs with overall vanishing color charge.
 Given the renormalization group equations (RGEs) satisfied by the hard functions and the transverse PDFs, it is straightforward to derive the ones for the soft functions. We find
 \begin{align}
 \label{RGE_sft}
   \frac{d}{d\ln\mu} \bm{\mathcal{S}}_{i\bar{i}}(\mu) = -\bm{\gamma}^{s\dagger}_{i\bar{i}}(\alpha_s) \, \bm{\mathcal{S}}_{i\bar{i}}(\mu) - \bm{\mathcal{S}}_{i\bar{i}}(\mu) \, \bm{\gamma}^{s}_{i\bar{i}}(\alpha_s) \, ,
 \end{align}
 with $\bm{\gamma}^s_{i\bar{i}} = \bm{\gamma}^h_{i\bar{i}} - 2\gamma^i \bm{1}$, where $\bm{\gamma}^h_{i\bar{i}}$ enter the RGEs of the hard functions and can be found in \cite{Ahrens:2010zv}.
 Following the approach shown in Ref.~\cite{BecherNeubert}, we can get $\bm{S}_{i\bar{i}}$ in Eq.~(\ref{eq:c}) from  Eq.~(\ref{RGE_sft}).

Given the resummed formula (\ref{eq:c}), it is important  to check whether its fixed-order expansion agrees with the exact results in the small $q_T$ region. To this end we expand Eq.~(\ref{eq:c}) to $\mathcal{O}(\alpha_s)$ and plug it into Eq.~(\ref{eq:master}). The results can be written as
\begin{align}
  \label{eq:nlo}
  &\frac{d^4\sigma}{dq_T^2 \, dy \, dM \, d\cos\theta} = \frac{\beta_t\alpha_s^3}{4sMq_T^2} \sum_{i} \frac{1}{d_i} \nonumber
  \\
  &\times \Bigg\{ f_{i/N_1}(\xi_1) \, f_{\bar{i}/N_2}(\xi_2) \, \mathrm{Tr} \left[ \bm{H}^{(0)}_{i\bar{i}} \left( \bm{A}_{i\bar{i}} \ln\frac{M^2}{q_T^2} + \bm{B}_{i\bar{i}} \right) \right] \nonumber
  \\
  &\quad + \mathrm{Tr} \left[ \bm{H}^{(0)}_{i\bar{i}} \bm{S}^{(0)}_{i\bar{i}} \right] \bigg[ \sum_{a} \big[ P_{ia}^{(1)} \otimes f_{a/N_1} \big](\xi_1) \, f_{\bar{i}/N_2}(\xi_2) \nonumber
  \\
  &\hspace{5em} + \sum_{b} f_{i/N_1}(\xi_1) \, \big[ P_{\bar{i}b}^{(1)} \otimes f_{b/N_2} \big](\xi_2) \bigg] \Bigg\} \, .
\end{align}
where
\begin{align}
  \label{eq:nloab}
  \bm{A}_{i\bar{i}} &= \Gamma^i_0 \, \bm{S}^{(0)}_{i\bar{i}} \, , \nn
  \\
  \bm{B}_{i\bar{i}} &= 2\gamma^{i}_0 \, \bm{S}^{(0)}_{i\bar{i}} - 4 \bm{w}^{33}_{i\bar{i}} + \frac{2(1+\beta_t^2)\ln x_s}{\beta_t} \, \bm{w}^{34}_{i\bar{i}} \nn
  \\
  & - 8 \ln\frac{-t_1}{m_t M} \, \bm{w}^{13}_{i\bar{i}} - 8 \ln\frac{-u_1}{m_t M} \, \bm{w}^{23}_{i\bar{i}} \, ,
\end{align}
and $\bm{H}^{(0)}_{i\bar{i}}$ is the LO hard function, which can be found, {\it e.g.}, in Eqs.~(61) and (62) of Ref.~\cite{Ahrens:2010zv}. The terms involving the $\bm{w}$ matrices originate from the soft function.  Eq.~(\ref{eq:nlo}) is also useful for $Q_T$ resummation in traditional CSS formalism.

\begin{figure}[h!]
  \includegraphics[scale=0.4]{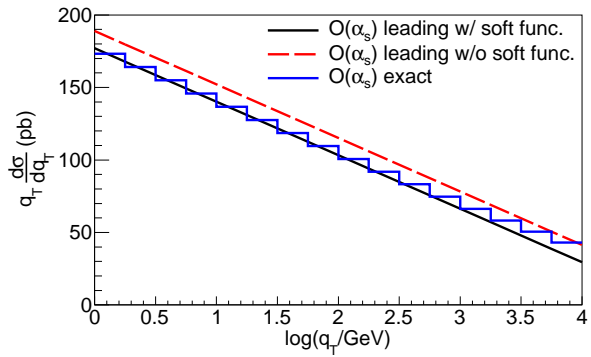}
  \vspace{-2ex}
  \caption{\label{fig:nlo} Comparison of the leading singular and the exact $\mathcal{O}(\alpha_s)$ distributions in the small $q_T$ region. Leading singular terms with (dashed-dotted line) and without (dashed line) the soft function contributions are presented.}
\end{figure}

Eq.~(\ref{eq:nlo}) captures the leading singular terms at order $\alpha_s$ in the limit $q_T \to 0$, which can be compared to the exact result in the small $q_T$ region. We show in Fig.~\ref{fig:nlo} the result from Eq.~(\ref{eq:nlo}) and the exact result calculated using \texttt{MCFM} \cite{Campbell:2000bg}. To illustrate the effect of the new soft functions, we also show in the plot the result without the contributions from the soft functions. As can be seen there, only when including the soft function contributions, the leading singular terms can reproduce the exact result, demonstrating the validity of our formalism. It is worth pointing out that our Eq.~(\ref{eq:nloab}) is in contradition with corresponding formulas in \cite{oldresum}.

 As a further check  of the  NLO soft functions, we employ the $Q_T$-subtraction method \cite{Catani:2007vq} to compute the NLO total cross
section for this process. We choose the MSTW2008NLO PDFs \cite{Martin:2009bu} and set
$m_t=172.5$~GeV.
Using the results presented in this Letter, we find that
the NLO total cross section is 161.93~pb and 162.30~pb, with and without the $L_\perp$-independent terms in the soft functions, respectively, while the result calculated by \texttt{MCFM}~\cite{Campbell:2000bg} is 161.94~pb. It turns out that the  numerical effects of $L_\perp$-independent terms are small, due to significant cancellation between initial-final contributions and final-final contributions.

To resum the large logarithms, we choose the default hard scale as $\mu_h=m_t$, and evolve the hard functions to a low scale $\mu$, where the soft functions and the transverse PDFs are evaluated.
  We follow the choice of $\mu$ proposed in \cite{BecherNeubert}, $\mu_i=q_i^* + q_T$ for $i=q,g$, where $q_i^*$ is determined by $q_i^*=M\exp(-2\pi/(\Gamma^i_0\als(q_i^*)))$. We also adopt the modified power counting such that $\als L^2_\perp$ is counted as $\mathcal{O}(1)$. Note that since $M \geq 2m_t \approx 345$~GeV, $q_q^* \gtrsim 3.0$~GeV which is considerably larger than that for $Z$-boson production, where $q^* \approx 1.88$~GeV. We therefore expect much weaker dependence on non-perturbative effects in $t\bar{t}$ production, down to $q_T=0$. Finally, to take into account the power corrections at large $q_T$, we match the NNLL resummed formula onto the exact NLO results~\cite{Campbell:2000bg}, and our best prediction is therefore of NLO+NNLL accuracy.

\begin{figure}[h!]
  \includegraphics[scale=0.4]{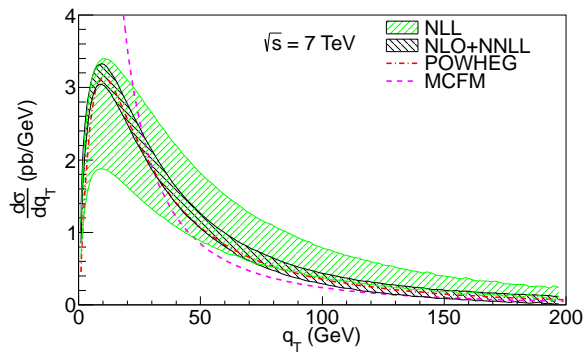}
  \vspace{-2ex}
  \caption{\label{fig:scale} Resummed predictions for the $q_T$ distribution at NLL (green band) and NLO+NNLL (black band). Also shown are the predictions of \texttt{POWHEG} and \texttt{MCFM}.}
\end{figure}

Fig.~\ref{fig:scale} shows the resummed $q_T$ distributions at the NLL and NLO+NNLL accuracy for $t\bar{t}$ production at the LHC with $\sqrt{s}=7$~TeV.
 Here and below we set $m_t=172.5$~GeV and use MSTW2008NNLO PDFs \cite{Martin:2009bu}.
 Uncertainties of the theoretical predictions are estimated by varying independently the common scale $\mu$ and the hard scale $\mu_h$ by a factor of two around their central values. It's clear from Fig.~\ref{fig:scale} that the NLO+NNLL prediction exhibits significantly smaller scale uncertainties, compared with the NLL one. As shown in Fig.~\ref{fig:scale}, the fixed-order prediction from \texttt{MCFM} is not reliable when $q_T$ is small, while the NLO+PS prediction of \texttt{POWHEG} \cite{Frixione:2007nw} is in good agreement with our NLO+NNLL resummed distribution. It should be noted that the \texttt{POWHEG} prediction exhibits a much larger scale dependence than the NLO+NNLL result, which is not shown in the plot.

\begin{figure}[h!]
  \includegraphics[scale=0.4]{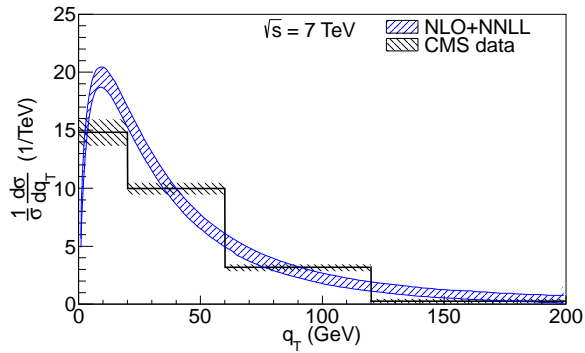}
  \vspace{-2ex}
  \caption{\label{fig:exp} Comparison of NLO+NNLL resummed prediction (blue band) for the normalized $q_T$ distribution with the experimental data from the CMS collaboration.}
\end{figure}

\begin{figure}[h!]
  \includegraphics[scale=0.4]{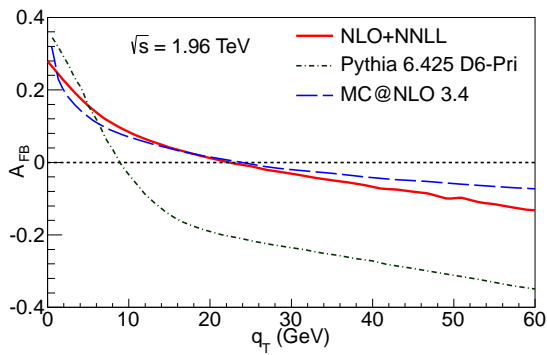}
  \vspace{-2ex}
  \caption{\label{fig:afb} The top quark charge asymmetry as a function of $q_T$. The \texttt{Pythia} and \texttt{MC@NLO} curves are extracted from \cite{Abazov:2011rq}.}
\end{figure}

In Fig.~\ref{fig:exp} we show our NLO+NNLL resummed prediction for the normalized $q_T$ distribution, together with the experimental data from the CMS collaboration \cite{TOP-11-013}, using an integrated luminosity of 1.14~fb$^{-1}$ at the LHC with $\sqrt{s}=7$~TeV.
 In this plot a non-perturbative factor of the form $\exp(-\Lambda_{\text{NP}}^2b^2)$ is included for the $q\bar{q}$-channel, with $\Lambda_{\text{NP}}=0.6$~GeV \cite{BecherNeubert}. For the $gg$-channel, the relevant scale is $q_g^* \gtrsim 14.0$~GeV, we therefore do not consider non-perturbative effects here.
The experimental data shows good agreement with our resummed prediction.

We finally turn to the $q_T$-dependent top quark charge asymmetry $A_{\rm FB}$. This quantity is of substantial interest because it will provide new hints for the puzzle of large deviation in $A_{\rm FB}$ observed at the Tevatron. In QCD, the asymmetry starts at NLO, however, it was found that an LO parton shower program like \texttt{Pythia} can exhibit non-zero $A_{\rm FB}$. As was explained in \cite{Skands:2012mm}, this is due to the fact that in the hard process $q\bar{q}\to t\bar{t}$, color coherence of the parton shower pushes the top-quark pair to higher transverse momentum when the top goes backwards. In our resummation formalism, this color coherence is accounted for by the soft function $\bm{S}_{q\bar{q}}$, whose dependence on $t_1$ and $u_1$ is asymmetric. In Fig.~\ref{fig:afb}, we present our resummed prediction for this observable, together with predictions from \texttt{MC@NLO} and \texttt{Pythia} extracted from \cite{Abazov:2011rq}. Interestingly, our NLO+NNLL resummed prediction shows very good agreement with the NLO+PS program \texttt{MC@NLO}. In particular, they predict the same cross-over at $q_T \sim 25$~GeV.

In conclusion, for the first time, we have presented a resummation framework for the transverse-momentum spectrum of top-quark pairs at hadron collider, valid up to arbitrary logarithmic accuracy. Compared with Drell-Yan and Higgs production, a new ingredient in our formalism is the introduction of the transverse soft function matrices, which describe the soft gluon effects associated with final-state radiations. We have explicitly shown that when expanded to $\mathcal{O}(\als)$, our resummation formula reproduces precisely the fixed-order prediction from \texttt{MCFM} at small $q_T$. We have carried out the resummation at NNLL accuracy. Our results agree quite well with those from parton shower programs and with the CMS measurement, while exhibiting a small scale dependence. We have also examined the $q_T$-dependent top quark charge asymmetry, which could help clarifying the large deviation from the SM observed at the Tevatron. Our formalism can also be applied to the $b\bar{b}$, $c\bar{c}$ production, as well as the production of colored supersymmetric partners. With the NNLO soft function which may be calculated in the future, our work provides a new subtraction method for computing the $t\bar{t}$ differential cross sections at NNLO, following the $q_T$ subtraction method of \cite{Catani:2007vq}. Finally, it is interesting to incorporate the decays of the top quark into our framework in a way similar to \cite{Melnikov:2011qx}, which we leave for future works.

This work was supported in part by the National Natural Science Foundation of China under Grants No. 11021092, No. 10975004 and No. 11135003, and by the Schweizer Nationalfonds under grant 200020-141360/1.

\end{document}